\pdfoutput=1

\documentclass[11pt,twoside,a4paper,cmspaper,final,collab]{cms-tdr}
\def\svnVersion{456238P}\def\cmsCernNoTag{CERN-EP-2017-331}\def\cmsCernDate{\today}\def\cmsMessage{Published in Physical Review D as \href{http://dx.doi.org/10.1103/PhysRevD.97.072010}{\doi{10.1103/PhysRevD.97.072010.}}}

\begin{document}\cmsNoteHeader{BPH-15-002}

\hyphenation{had-ron-i-za-tion}
\hyphenation{cal-or-i-me-ter}
\hyphenation{de-vices}
\RCS$HeadURL: svn+ssh://svn.cern.ch/reps/tdr2/papers/BPH-15-002/trunk/BPH-15-002.tex $
\RCS$Id: BPH-15-002.tex 456204 2018-04-18 19:52:21Z rreyesal $

\newlength\cmsTabSkip\setlength{\cmsTabSkip}{1ex}

\newcommand{\ppi}      {\ensuremath{\Pp\Pgpm}\xspace}
\newcommand{\pipi}      {\ensuremath{\Pgpp\Pgpm}\xspace}

\newcommand{\Lambdab}	{\ensuremath{\PgL_{\cmsSymbolFace{b}}}\xspace}
\newcommand{\aLambdab}	{\ensuremath{\PagL_{\cmsSymbolFace{b}}}\xspace}
 \newcommand{\PGSb}	{\ensuremath{\Sigma_{\cmsSymbolFace{b}} }\xspace}
 \newcommand{\PGSbsta}	{\ensuremath{\Sigma^{\ast}_{\cmsSymbolFace{b}} }\xspace}
 \newcommand{\PGSbpm}{\ensuremath{\PGSb^{\pm}}\xspace}
 \newcommand{\PGSbstapm}{\ensuremath{\PGSb^{\ast\pm}}\xspace}

\newcommand{\LbJpsiL}   {\ensuremath{\Lambdab\to\PJGy\PgL}\xspace}
\newcommand{\aLbJpsiL}   {\ensuremath{\aLambdab\to\PJGy\PagL}\xspace}
\newcommand{\Lbdecay}   {\ensuremath{\Lambdab\to\PJGy(\to\MM)\PgL(\to\ppi)}\xspace}

\newcommand{\JMM}   {\ensuremath{\PJGy\to\MM}\xspace}
\newcommand{\ptmu}      {\ensuremath{\pt^{\Pgm}}\xspace}
\newcommand{\ptpi}      {\ensuremath{\pt^{\Pgp}}\xspace}
\newcommand{\ptp}      {\ensuremath{\pt^{\Pp}}\xspace}
\newcommand{\ptJL}      {\ensuremath{\pt^{\PJGy\PgL}}\xspace}

\newcommand{\MLb}	{\ensuremath{m_{\PJGy\PgL}}\xspace}
\newcommand{\aMLb}	{\ensuremath{m_{\PJGy\PagL}}\xspace}

\newcommand{\thetaL}    {\ensuremath{\theta_{\PgL}}\xspace}
\newcommand{\Omegap}    {\ensuremath{\Omega_{\Pp}}\xspace}
\newcommand{\thetap}    {\ensuremath{\theta_{\Pp}}\xspace}
\newcommand{\phip}    {\ensuremath{\varphi_{\Pp}}\xspace}
\newcommand{\Omegam}    {\ensuremath{\Omega_{\Pgm}}\xspace}
\newcommand{\thetam}    {\ensuremath{\theta_{\Pgm}}\xspace}
\newcommand{\phim}    {\ensuremath{\varphi_{\Pgm}}\xspace}

\newlength\cmsFigWidth
\ifthenelse{\boolean{cms@external}}{\setlength\cmsFigWidth{0.85\columnwidth}}{\setlength\cmsFigWidth{0.4\textwidth}}
\ifthenelse{\boolean{cms@external}}{\providecommand{\cmsLeft}{top\xspace}}{\providecommand{\cmsLeft}{left\xspace}}
\ifthenelse{\boolean{cms@external}}{\providecommand{\cmsRight}{bottom\xspace}}{\providecommand{\cmsRight}{right\xspace}}
\cmsNoteHeader{BPH-15-002}
\title{\texorpdfstring{Measurement of the \Lambdab polarization and angular parameters in \LbJpsiL decays from {\Pp\Pp} collisions at  $\sqrt{s}= 7$ and 8\TeV}{Measurement of the Lambda[b] polarization and angular parameters in Lambda[b] to J/psi Lambda decays from pp collisions at  sqrt(s) = 7 and 8 TeV}}

\date{\today}
\abstract{
An analysis of the bottom baryon decay \Lbdecay is performed to measure the \Lambdab polarization and three angular parameters in data from {\Pp\Pp} collisions at $\sqrt{s} = 7$ and 8\TeV, collected by the CMS experiment at the LHC. The \Lambdab polarization is measured to be $0.00 \pm 0.06\stat\pm 0.06\syst$  and the parity-violating asymmetry parameter is determined to be $0.14 \pm 0.14\stat \pm 0.10\syst$. The measurements are compared to various theoretical predictions, including those from perturbative quantum chromodynamics.
}

\hypersetup{
pdfauthor={CMS Collaboration},
pdftitle={Measurement of the Lambda[b] polarization and angular parameters in Lambda[b] to Lambda J/psi  decays from pp collisions at sqrt(s)= 7 and 8 TeV},
pdfsubject={CMS},
pdfkeywords={CMS,b physics, baryon decay, polarization}}

\maketitle
\section{Introduction}
The decay \LbJpsiL is a rich source of information about the effect of strong interactions in hadronic decays.  For this particular decay, perturbative quantum chromodynamics can be applied and therefore a systematic approach can be taken to study its characteristics.  Several techniques~\cite{A2, A3,A4,LEITNER2005435,A7, A8,angdist,kramer,Gutsche,A18} are used to study and calculate the decay amplitudes and the effect of the $\cPqb{\>}$quark polarization on this decay.  The most interesting parameters that can be measured are the polarization, $P$, and the parity-violating decay asymmetry of the \Lambdab, which is called $\alpha_b$ in some papers and is equal to $-\alpha_1$ in the notation used in this analysis. The LHCb and ATLAS experiments  have reported measurements on this decay~\cite{LHCb, ATLAS}. The LHCb Collaboration measured the \Lambdab polarization and the decay amplitudes,  while ATLAS assumed a \Lambdab polarization of zero and measured the amplitudes. In this paper, a measurement of the  \Lambdab transverse polarization is presented using the decay \LbJpsiL, with \JMM  and $\PgL\to \ppi$. Charge-conjugate modes are implied throughout this paper unless otherwise stated. The \Lambdab baryons used in this measurement come from both direct production in {\Pp\Pp} collisions and decays of heavier {\cPqb} baryons~\cite{A2,A9,A10,Aaltonen:2013tta,A11}.
The data were collected with the CMS detector in {\Pp\Pp} collisions at center-of-mass energies of 7 and 8\TeV, corresponding to integrated luminosities of 5.2 and 19.8\fbinv, respectively.

\section{Angular distribution}
\begin{figure}[!ht]
        \centering
                \includegraphics[width=0.49\textwidth]{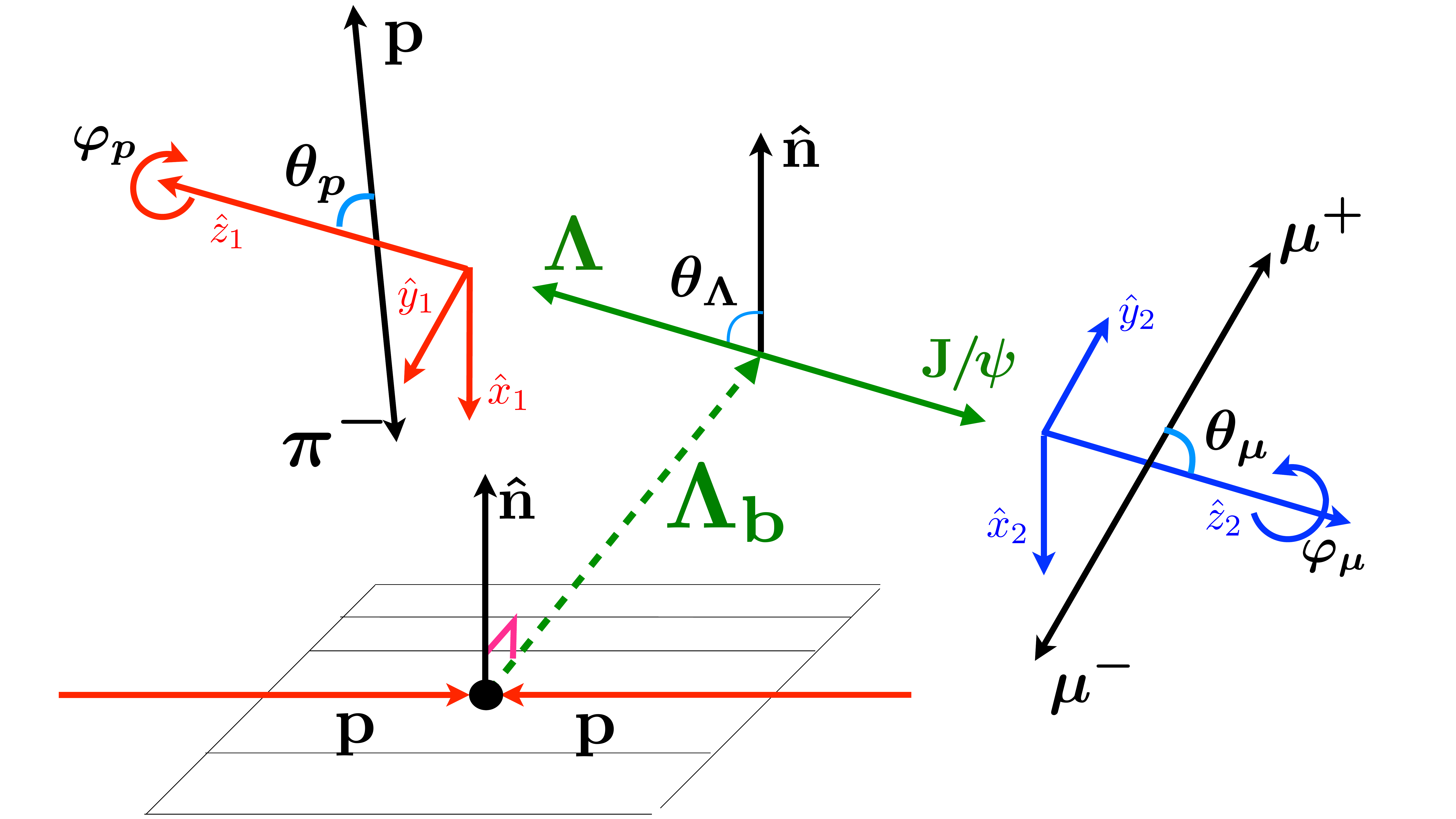}
                \caption{Definition of the angles used to describe the \LbJpsiL decay into the \MM\ppi final state as explained in the text.}
                \label{fig:angleDiagram}
\end{figure}

The \LbJpsiL decay into the \MM\ppi final state is illustrated in Fig.~\ref{fig:angleDiagram}. In {\Pp\Pp} collisions, we define the polarization of the \Lambdab as the mean value of the \Lambdab spin along the unit vector:
\begin{equation}
\centering
\hat{n}=\frac{ \vec{p}_{\text{beam}}\times\vec{p}_{\Lambdab}}{ \abs{\vec{p}_{\text{beam}}\times\vec{p}_{\Lambdab}}},
\end{equation}
normal to its production plane, where $\vec{p}_{\text{beam}}$ is in the direction of the counterclockwise proton beam direction~\cite{Chatrchyan:2008zzk}, and  $\vec{p}_{\Lambdab}$ is the \Lambdab momentum. The decay is described by four complex helicity amplitudes $T_{\lambda_1 \lambda_2}$, with $\lambda_1 = \pm1/2$ and $\lambda_2 = \pm1, 0$ referring to the helicities of the {\PgL} and \PJGy particles, respectively.  The angular distribution is a function of five decay angles $(\thetaL, \thetap, \thetam, \phip, \phim)$  and has the form~\cite{kramer}
\ifthenelse{\boolean{cms@external}}{
\begin{multline}
\frac{\rd^{5}\Gamma}{\rd \cos\thetaL \rd\Omegap \rd\Omegam}\left(\thetaL, \thetap, \thetam, \phip, \phim \right) \\= \frac{1}{(4\pi)^{3}}\sum_{i=1}^{20}u_{i}(T_{\lambda_1 \lambda_2})\,v_{i}(P,\alpha_{\PgL})\,w_i(\thetaL, \thetap, \thetam, \phip, \phim),
\end{multline}
}{
\begin{equation}
\frac{\rd^{5}\Gamma}{\rd\cos\thetaL \rd\Omegap \rd\Omegam}\left(\thetaL, \thetap, \thetam, \phip, \phim \right) = \frac{1}{(4\pi)^{3}}\sum_{i=1}^{20}u_{i}(T_{\lambda_1 \lambda_2})\,v_{i}(P,\alpha_{\PgL})\,w_i(\thetaL, \thetap, \thetam, \phip, \phim),
\end{equation}
}
 where $w_i $ are trigonometric functions, $u_i$ are bilinear combinations of the helicity amplitudes $T_{\lambda_1 \lambda_2}$, and $v_i$ stands for  1, $P$, $\alpha_{\PgL}$, or $P\alpha_{\PgL}$; $P$ is the \Lambdab polarization and $\alpha_{\PgL}$ is the asymmetry parameter in the decay $\PgL\to \ppi$. The angle $\thetaL$ is the polar angle of the {\PgL} momentum relative to $\hat{n}$ in the \Lambdab rest frame; $\thetap$ and $\phip$ are the polar and azimuthal angles of the proton, respectively,  defined with respect to the axes $\hat{z}_1=\vec{p}_{\PgL}/\abs{\vec{p}_{\PgL}}$
and $\hat{y}_{1}=\left( \hat{n} \times \vec{p}_{\PgL} \right)/\abs{\hat{n} \times \vec{p}_{\PgL}}$ in the rest frame of the \PgL; and the angles $\thetam$ and $\phim$ are the polar and azimuthal angles, respectively, of the positively charged muon, defined with respect to the axes $\hat{z}_2=\vec{p}_{\PJGy}/\abs{\vec{p}_{\PJGy}}$ and $\hat{y}_2=\left( \hat{n} \times \vec{p}_{\PJGy}\right)/\abs{\hat{n} \times \vec{p}_{\PJGy}}$ in the \PJGy rest frame. Here, $\vec{p}_{\PgL}$ and $\vec{p}_{\PJGy}$ are the momenta of the {\PgL} and \PJGy, respectively, and $\rd\Omegap=\rd(\cos\thetap)\rd\phip$ and $\rd\Omegam=\rd(\cos\thetam)\rd\phim$ are  differential solid angles.
Assuming uniform detector acceptance over the azimuthal angles \phip and \phim, the angular distribution can be simplified through an integration over these two  angles:
\ifthenelse{\boolean{cms@external}}{
\begin{multline}\label{eq:AngDist}
\frac{\rd^{3}\Gamma}{\rd\cos\thetaL\,\rd\cos\thetap\,\rd\cos\thetam}(\thetaL,\thetap,\thetam) \\=
 \int_{-\pi}^{\pi}\int_{-\pi}^{\pi}\frac{\rd^{5}\Gamma}{\rd\cos\thetaL\,\rd\Omegap\,\rd\Omegam}(\thetaL,\thetap,\thetam,\phip,\phim)\rd\phip\,\rd\phim \\
  \sim\sum_{i=1}^{8}u_{i}\left(\abs{T_{\lambda_1 \lambda_2}}^{2}\right)v_{i}(P,\alpha_{\PgL})w_{i}(\thetaL,\thetap,\thetam).
\end{multline}
}{
\begin{equation}\label{eq:AngDist}
\begin{aligned}
\frac{\rd^{3}\Gamma}{\rd\cos{\thetaL}\,\rd\cos{\thetap}\,\rd\cos{\thetam}}(\thetaL,\thetap,\thetam) &=
 \int_{-\pi}^{\pi}\int_{-\pi}^{\pi}\frac{\rd^{5}\Gamma}{\rd\cos{\thetaL}\,\rd\Omegap\,\rd\Omegam}(\thetaL,\thetap,\thetam,\phip,\phim)\rd\phip\,\rd\phim \\
  &\sim\sum_{i=1}^{8}u_{i}\left(\abs{T_{\lambda_1 \lambda_2}}^{2}\right)v_{i}(P,\alpha_{\PgL})w_{i}(\thetaL,\thetap,\thetam).
  \end{aligned}
\end{equation}
}
The eight functional forms of $u_i$, $v_i$, and $w_i$  are listed in Table~\ref{table:functions}. The $u_{i}$ factors  are written in terms of the three angular decay parameters $\alpha_1$, $\alpha_2$, and $\gamma_0$ proposed in Ref.~\cite{kramer}, and the constant~1, which themselves can be written in terms of the $T_{\lambda_1 \lambda_2}$  amplitudes as
\begin{equation}\begin{aligned}\label{eq:DecayParam}
1&=  \abs{ T_{++}}^{2}+\abs{ T_{+0}}^{2}+\abs{ T_{-0}}^{2}+\abs{ T_{--}}^{2},\\
\alpha_{1}&=  \abs{ T_{++}}^{2}-\abs{ T_{+0}}^{2}+\abs{ T_{-0}}^{2}-\abs{ T_{--}}^{2},\\
\alpha_{2}&=  \abs{ T_{++}}^{2}+\abs{ T_{+0}}^{2}-\abs{ T_{-0}}^{2}-\abs{ T_{--}}^{2},\\
\gamma_{0}&= \abs{ T_{++}}^{2}-2\abs{ T_{+0}}^{2}-2\abs{ T_{-0}}^{2}+\abs{ T_{--}}^{2},
\end{aligned}\end{equation}
where $\alpha_1$ is the asymmetry parameter for the decay \LbJpsiL, $\alpha_{2}$ represents  the longitudinal polarization of the \PgL, and $\gamma_{0}$ is a parameter that depends on the longitudinal and transverse polarizations of the {\PJGy}\cite{Gutsche}. The CP invariance of Eq.~(\ref{eq:AngDist})  implies that the parameters for \Lambdab and \aLambdab are related as follows:
\begin{equation}\begin{aligned}\label{eq:decayParamRel}
\overline{P} 	&=  -P,\\
\overline{\alpha}_1 &= -\alpha_1,\\
\overline{\alpha}_2 &=  -\alpha_2,\\
\overline{\gamma}_0 &=   \gamma_0.
\end{aligned}\end{equation}
In addition, CP conservation in $\PgL\to \ppi$ decays implies that $\overline{\alpha}_{\PgL} =  -\alpha_{\PgL}$~\cite{pdg}.
In this analysis, the four parameters $(P, \alpha_1, \alpha_2, \gamma_0)$ are extracted from an analysis of the angular distribution given in Eq.~(\ref{eq:AngDist}), where $\alpha_{\PgL}$ is fixed to its world-average  value of $0.642 \pm 0.013$~\cite{pdg}.
\begin{table*}[ht!]
\centering
\topcaption{Functions used in Eq.~(\ref{eq:AngDist}) to describe the angular distribution in the decay \LbJpsiL, with \JMM and $\PgL\to \ppi$.}
\label{table:functions}
\begin{scotch}{l l l l}
$i$ & $u_i$ & $v_i$ & $w_i$ \\ \hline
1 & 1 & 1 & 1 \\
2 & $\alpha_2$ & $\alpha_{\PgL}$ & $\cos\thetap$ \\
3 & $-\alpha_1$ & $P$ & $\cos\thetaL$ \\
4 & $-\left(1+2\gamma_{0}\right)/3$ & $\alpha_{\PgL}P$ & $\cos\thetaL\cos\thetap$ \\
5 & $\gamma_{0}/2$ & 1 & $\left(3\cos^{2}\thetam-1\right)/2$ \\
6 & $\left(3\alpha_{1}-\alpha_{2}\right)/4$ & $\alpha_{\PgL}$ & $ \cos\thetap\left(3\cos^{2}\thetam-1\right)/2$ \\
7 & $\left(\alpha_{1}-3\alpha_{2}\right)/4$ & $P$ & $ \cos\thetaL\left(3\cos^{2}\thetam-1\right)/2$ \\
8 & $\left(\gamma_{0}-4\right)/6$ & $\alpha_{\PgL}P$ & $ \cos\thetaL\cos\thetap\left(3\cos^{2}\thetam-1\right)/2$ \\
\end{scotch}
\end{table*}
\section{The CMS detector}
The CMS detector is used to study a wide range of phenomena produced in high-energy collisions, with its central feature being a superconducting solenoid of 6\unit{m} internal diameter, providing a magnetic field of  3.8\unit{T}. A silicon pixel and strip tracker, a lead tungstate scintillating crystal electromagnetic calorimeter, and a brass and scintillator sampling hadron calorimeter, including a central barrel and endcap detectors, are located within the magnetic volume.

The silicon tracker detects charged particles within the pseudorapidity range $\abs{\eta} < 2.5$. It consists of 1440 silicon pixel and 15\,148 silicon strip detector modules. For nonisolated particles with transverse momentum of $1 < \pt < 10\GeV$ and $\abs{\eta} < 1.4$, the track resolutions are typically 1.5\% in \pt and 25--90 (45--150)\mum in the transverse (longitudinal) impact parameter~\cite{TRK-11-001}. Muons are detected in gas-ionization chambers within the pseudorapidity range $\abs{\eta} < 2.4$, with detection planes made using three technologies: drift tubes, cathode strip chambers, and resistive plate chambers~\cite{Chatrchyan:2012xi}. The global event reconstruction (also called particle-flow event reconstruction~\cite{CMS-PRF-14-001}) consists of reconstructing and identifying each individual particle with an optimized combination of all subdetector information. In this process, muons are identified as a track in the silicon tracker consistent with either a track or several hits in the muon system, associated with an energy deficit in the calorimeters.

Events of interest are selected using a two-tiered trigger system~\cite{Khachatryan:2016bia}. The first level (L1), composed of custom hardware processors, uses information from the calorimeters and muon detectors to select events at a rate of around 100\unit{kHz} within a time interval of less than 4\mus. The second level, known as the high-level trigger (HLT), consists of a farm of processors running a version of the full event reconstruction software optimized for fast processing, and reduces the event rate to around 1\unit{kHz} before data storage.

A more detailed description of the CMS detector, a definition of the coordinate system used and the relevant kinematic variables, can be found in Ref.~\cite{Chatrchyan:2008zzk}.

\section{Data and simulated events}

We use data collected with a trigger designed for events containing a \PJGy meson decaying to two muons that form a displaced vertex relative to the mean {\Pp\Pp} collision point (beamspot). The requirement on the displacement does not affect the angular distributions of the reconstructed \Lambdab decay products. The dimuon trigger configurations were changed during the data taking at different center-of-mass energies, with increasingly stringent requirements to maintain an acceptable trigger rate as the instantaneous luminosity increased. The requirements of the different trigger versions are as follows: the \PJGy candidates are selected in the invariant mass window 2.5--4.0\GeV and 2.9--3.3\GeV depending on the version; the angle ($\beta$) between the reconstructed momentum vector of the dimuon system and the vector pointing from the beamspot position to the dimuon vertex must have a value of $\cos\beta > 0.9$; the distance between the beamspot and the dimuon vertex in the transverse plane must have a value that is at least a factor of three larger than its uncertainty (standard deviation  or SD); the muon pair must satisfy $\pt^{\Pgm\Pgm}>6.5$ or $6.9$\GeV  in the different versions; the $\chi^{2}$ probability of the fit of the two muons to a common vertex must exceed 0.05, 0.10, or 0.15 from  the earliest to the latest version; each muon must be in $\abs{\eta(\Pgm)} < 2.2$ and have $\ptmu >3.5$ or 4\GeV; and the distance of closest approach of each muon to the common vertex in the transverse plane must be less than 0.5\cm.

Simulated events of the signal decay are used to study the effects of detector acceptance and  selection on the reconstructed angular distributions.
The events are generated using \PYTHIA 6.4~\cite{pythia}  for production and hadronization, and \EVTGEN~\cite{evt} is used to describe the {\cPqb} and {\cPqc} hadron decays. The generated events are passed through the full CMS detector simulation based on \GEANTfour~\cite{geant4}. The simulated event samples are generated to reproduce $\sqrt{s}= 7$ and 8\TeV data-taking conditions, where additional simulation of $\Pp\Pp$ interactions in the same or nearby beam crossings  and the impact of the HLT are included.
Simulated events are reconstructed and selected using the same algorithms and requirements as used for data.

\section{Reconstruction and event selection}\label{section:reconstruction}

The offline selection requires  pairs of oppositely charged muons originating from a common vertex to form the \PJGy candidates. The standard CMS muon reconstruction procedure~\cite{Chatrchyan:2012xi} is used to identify the muons. Since the trigger changed slightly over the different data-taking periods, the offline selection is required to be more restrictive than the most-stringent trigger, and is summarized as follows: (i) each muon must have $\ptmu >4$\GeV  and the dimuon transverse momentum must satisfy $\pt^{\Pgm\Pgm}>8$\GeV; (ii) the $\chi^{2}$ probability must exceed 0.15; and (iii) the dimuon invariant mass must lie within $\pm$150\MeV of the world-average \PJGy mass~\cite{pdg}. Additional requirements are the same as the trigger selection and, to reduce background, the \PJGy candidates must satisfy $\cos\beta>0.99$.

The {\PgL}  candidates are constructed from pairs of oppositely charged tracks that have a successful fit to a common vertex. Since the default CMS algorithms cannot distinguish between pions and protons, the higher- and lower-momentum tracks are assumed to have the proton and pion  masses~\cite{pdg}, respectively. The selections used for {\PgL}  and {\PKzS} particles are detailed in Ref.~\cite{v0}. They are optimized to reduce background using the following additional requirements: (i) each track is required to have at least 6 hits in the silicon tracker and a $\chi^2$ track fit per degree of freedom $<7$; (ii) the tracks coming from the {\PgL} decay are required to have $\ptpi>0.3$\GeV,  $\ptp >1.0$\GeV; (iii) the transverse impact parameter of the tracks relative to the beamspot is required to be greater than 3 SD; (iv) the probability of the two-track vertex fit must exceed 2\%; (v) the transverse separation of the two-track vertex from the beamspot is required to be larger than 15 SD; (vi) the invariant mass of the {\PgL} candidate is selected to lie within $\pm9\MeV$ of the world-average value~\cite{pdg} and satisfy  $\pt^{\Pp\Pgp}>1.3$\GeV; and (vii) to reduce the contamination of $\PKzS \to \Pgpp \Pgpm$ decays, events are removed  if their invariant mass falls within $\pm20\MeV$ of the {\PKzS} mass when the proton candidate is given the charged pion mass.

The \Lambdab candidates are fitted to a common vertex by combining the \PJGy and {\PgL} candidates, with the respective mass constraints to the world-average values of the \PJGy and {\PgL} masses~\cite{pdg}. The selection of \Lambdab candidates is optimized to reduce background with the additional requirements: $\ptJL >10$\GeV, a $\chi^{2}$ probability of the fit to the $\PJGy\PgL$ vertex $ > 3\%$, and the $\PJGy\PgL$ invariant mass $5.40 < \MLb < 5.84$\GeV.

To extract the number of signal and background events and to define the signal and sidebands regions, unbinned maximum likelihood fits to the reconstructed invariant mass (\MLb) distributions are performed, using separate data sets of the \Lambdab and \aLambdab candidates at $\sqrt{s}= 7$ and 8\TeV. The signal shape is modeled by two Gaussian functions with different SDs, $\sigma_1$ and $\sigma_2$, but common mean $\mu_{\PJGy\PgL}$, and the background by  a first-order polynomial. We define in the four data sets the signal region as  $\mu_{\PJGy\PgL}\pm16\MeV$, the lower sideband region as $[5.46, 5.54]$\GeV, and the upper sideband region as $[5.70, 5.78]$\GeV.  From the fits the \Lambdab yields are $981\pm39$ and $2072\pm55$ signal events, and the \aLambdab yields are $916\pm40$ and $1974\pm53$ signal events at $\sqrt{s}= 7$ and 8\TeV, respectively.

\section{Measurement of the polarization and angular parameters}\label{sect:Measurement}

The analysis extracts the \Lambdab polarization, $P$, and the angular decay parameters $\alpha_1$, $\alpha_2$, and $\gamma_0$. The results  are obtained from an unbinned maximum likelihood fit to the $\PJGy\PgL$ invariant mass distribution and the three angular variables $\Theta_{3}=\left(\cos\thetaL,\cos\thetap,\cos\thetam\right)$, using the extended likelihood function:
\begin{equation}\label{eq:Flikelihood}
L=\text{exp}\left(-N_{\text{sig}}-N_{\text{bkg}}\right)\prod^{\text{N}}\left[N_{\text{sig}}\,PDF_{\text{sig}}+N_{\text{bkg}}\,PDF_{\text{bkg}}\right],
\end{equation}
where N is the total number of events, $N_{\text{sig}}$ and $N_{\text{bkg}}$ are the yields of signal and background events, respectively, determined from the fit in Sec.~\ref{section:reconstruction}, and $PDF_{\text{sig}}$ and $PDF_{\text{bkg}}$ represent the probability density functions ($PDFs$) for the signal and background, respectively. The $PDF_{\text{sig}}$  has the form
\begin{equation}
PDF_{\text{sig}}=F_{\text{sig}}(\Theta_{3})\,\epsilon(\Theta_{3})\,G(\MLb),
\end{equation}
where $F_{\text{sig}}$ represents the theoretical angular distribution given by Eq.~(\ref{eq:AngDist}) and \textit{G} is the sum of two Gaussian functions used to model the $\PJGy\PgL$ invariant mass distribution, as mentioned in Sec.\ref{section:reconstruction}. The effect of the detector on the angular distribution is taken into account by the factor $\epsilon$ that represents the efficiency as a function of the angles.

\begin{figure*}[htbp!]
        \centering
         \includegraphics[width=0.3\textwidth]{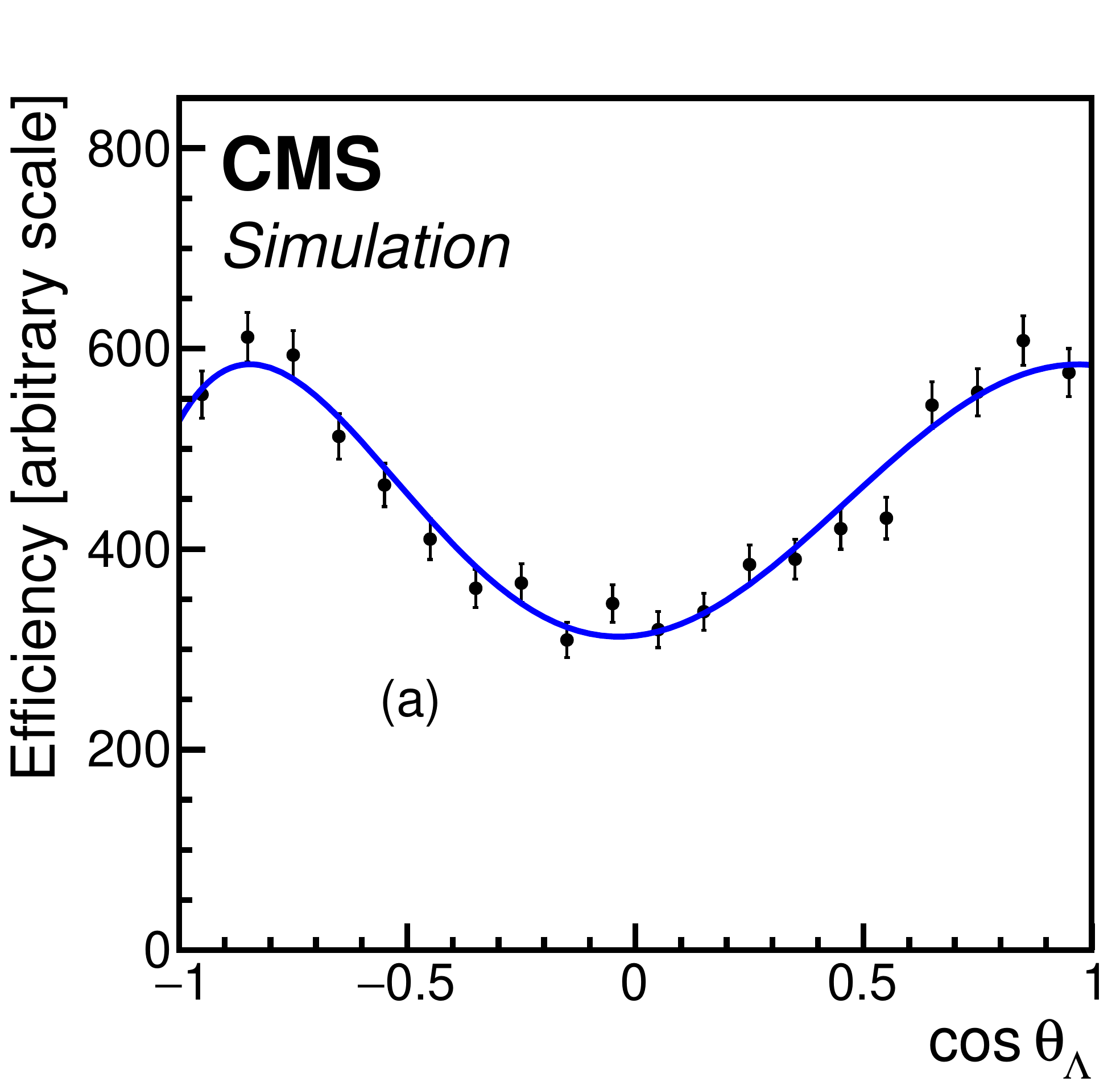}
         \includegraphics[width=0.3\textwidth]{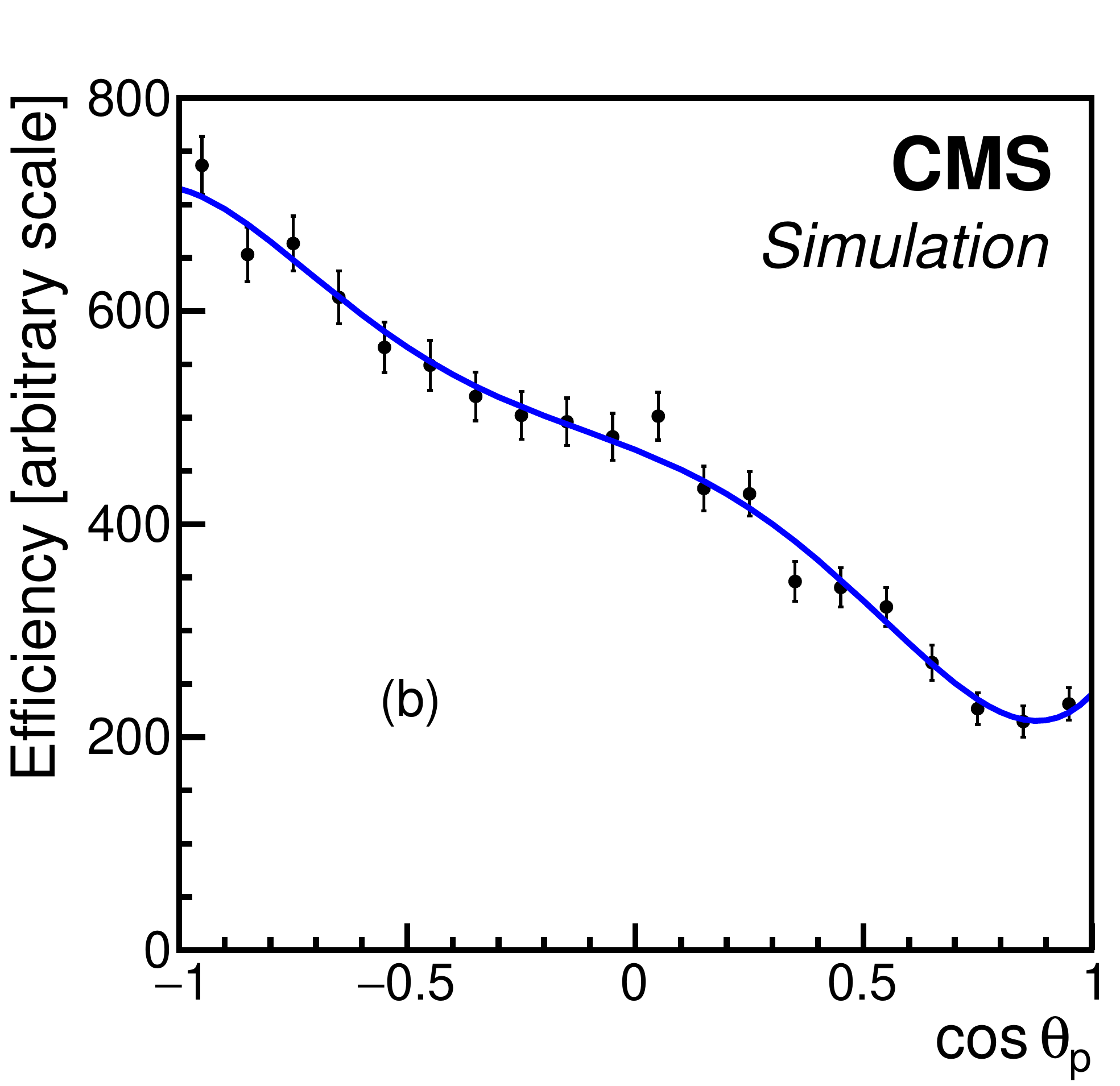}
         \includegraphics[width=0.3\textwidth]{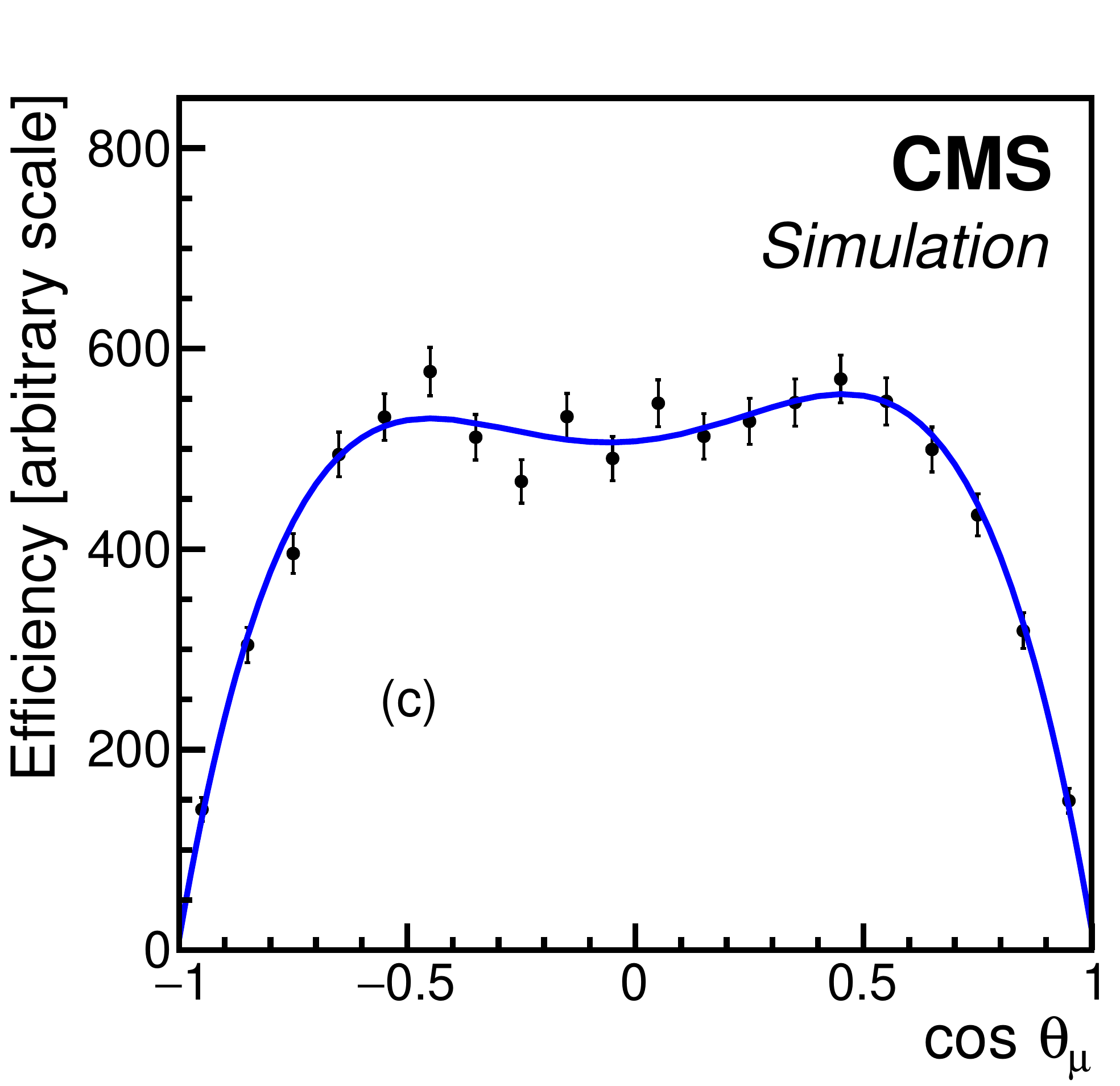}
\caption{The efficiencies as a function of (a) $\cos\thetaL$, (b) $\cos\thetap$, and  (c) $\cos\thetam$ obtained from simulated \LbJpsiL decays at $\sqrt{s}= 8$\TeV. The vertical bars on the points are the statistical uncertainties in the simulated data, and the lines show the projections of a 3D fit to the distributions using Chebyshev polynomials. The scales of the vertical axes are arbitrary.}\label{fig:Fitefficiency}
\end{figure*}

 To estimate the angular efficiency, simulated events of \LbJpsiL decays are generated with uniform distributions in $\cos\thetaL$, $\cos\thetap$, and $\cos\thetam$. After full detector simulation, reconstruction, and implementation of the final selection requirements, the slight differences between the simulated events and the background-subtracted data  are minimized through a weighting procedure where weights are assigned to the simulated events to match the data. The weights are computed with an iterative process in which, for each iteration, the  histograms of a selection variable in background-subtracted data and simulated events are used to calculate the ratio bin by bin (weight) with its propagated statistical uncertainty. The final weight per event is the product of the weights in each iteration. The efficiency distributions as a function of the variables are fit with a product of Chebyshev polynomials, where the coefficients are obtained for \Lambdab and \aLambdab at  $\sqrt{s}= 7$ and 8\TeV in separate likelihood fits. The simulated efficiency distributions and the results of these fits are shown in Fig.~\ref{fig:Fitefficiency}  for the \Lambdab candidates at $\sqrt{s} = 8$\TeV.

\begin{figure*}[htbp!]
               \centering
                \includegraphics[width=0.3\textwidth]{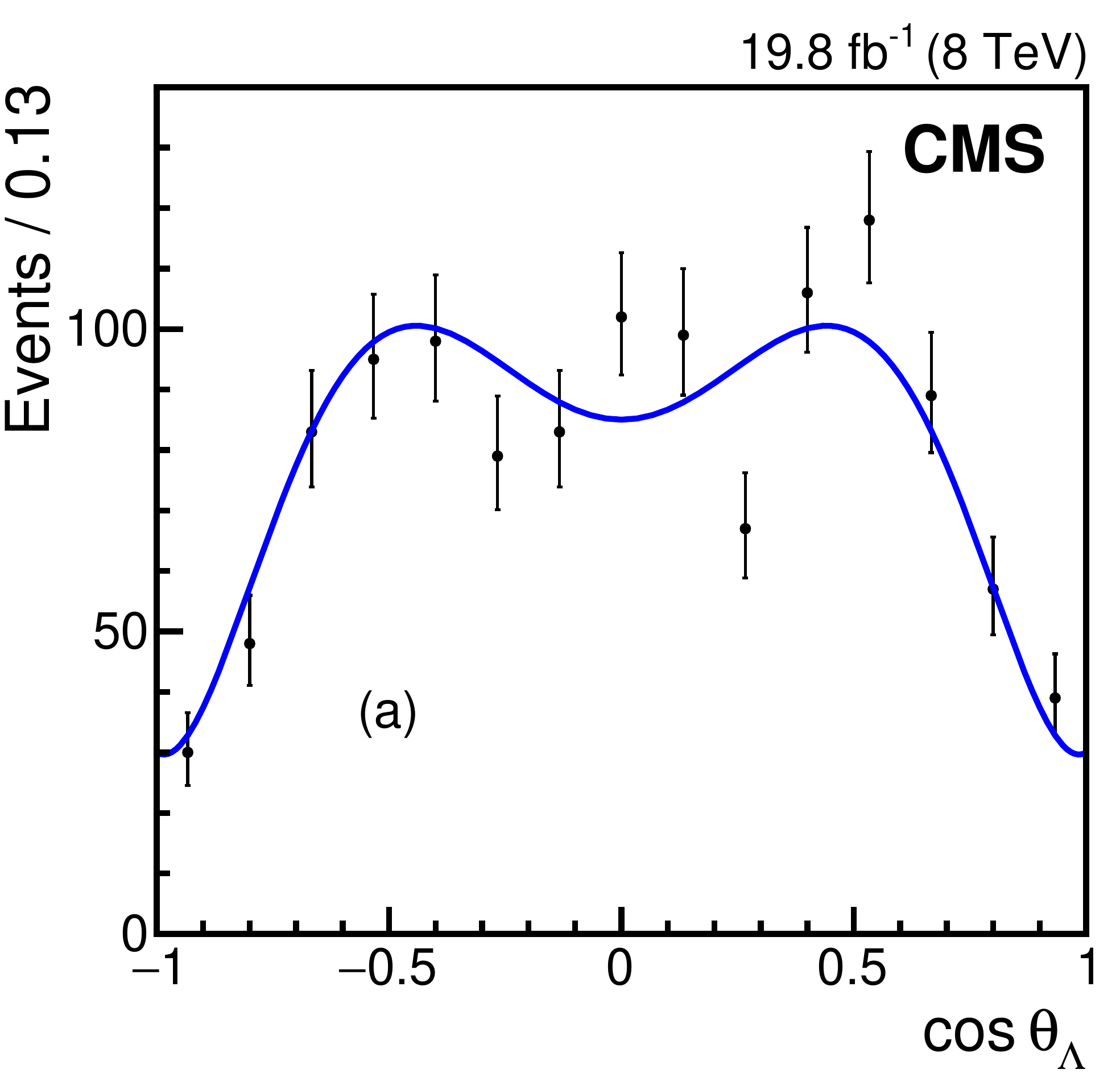}
                \includegraphics[width=0.3\textwidth]{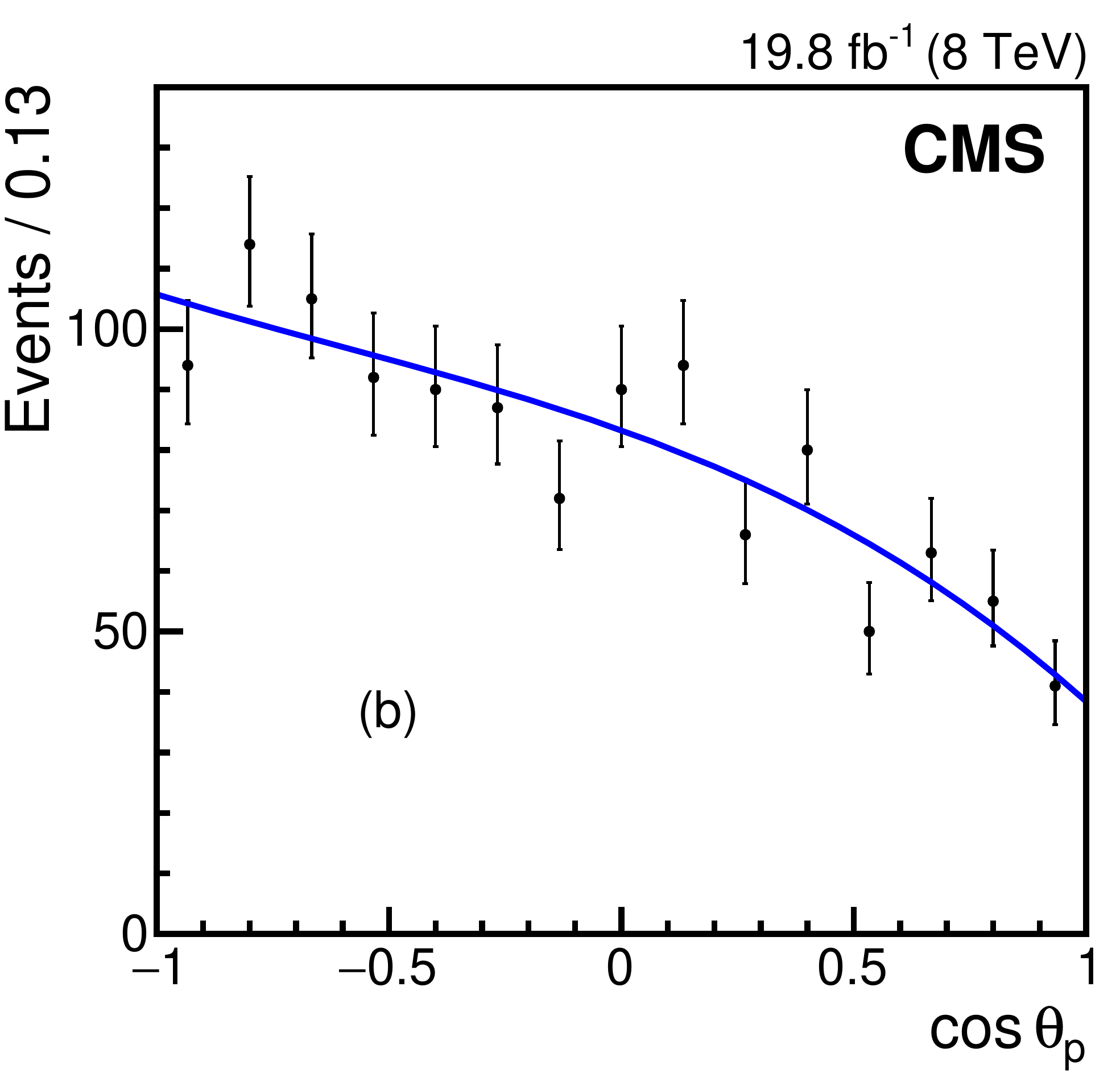}
                \includegraphics[width=0.3\textwidth]{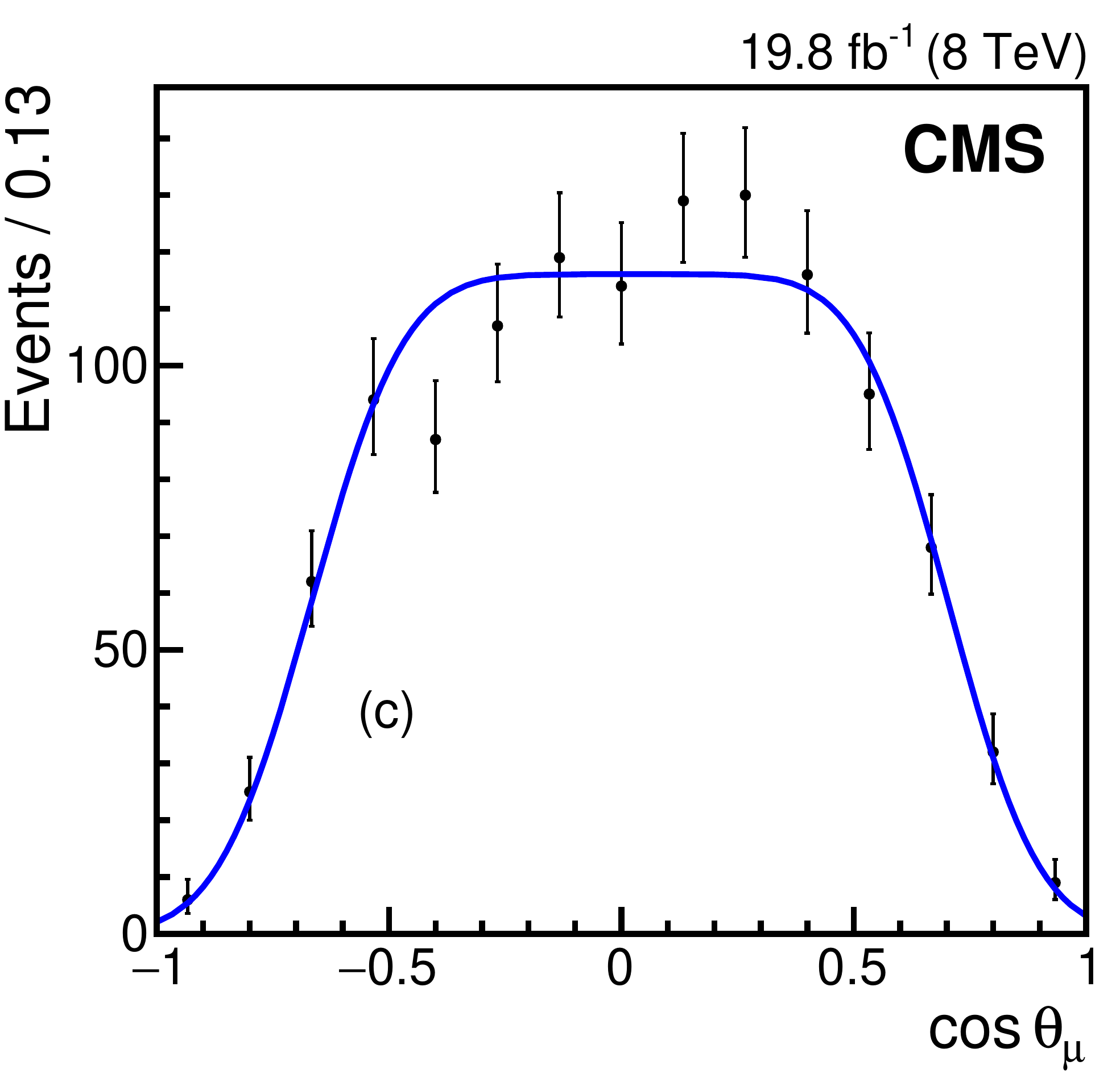}
 \caption{The background angular distributions of (a) $\cos\thetaL$, (b) $\cos\thetap$, and (c) $\cos\thetam$ are shown, as obtained from the sidebands in the $\PJGy\PgL$ invariant mass distribution at $\sqrt{s}= 8$\TeV. The vertical bars on the points represent the statistical uncertainties, and the solid lines are the results of the fits to data as described in the text.}
         \label{fig:sidebands}
\end{figure*}

 The background $PDF_\text{bkg}$ is given by the product of a first-order polynomial $\mathcal{P}(\MLb)$ for the invariant mass and an angular distribution function $F_{\text{bkg}}\left(\Theta_{3}\right)$:
\begin{equation}
PDF_{\text{bkg}}=\mathcal{P}(\MLb)\,F_{\text{bkg}}\left(\Theta_{3}\right).
\end{equation}
The  background angular distributions $F_{\text{bkg}}\left(\Theta_{3}\right)$ are estimated using events from the \MLb invariant mass sidebands. They are modeled using Chebyshev polynomials for $\cos\thetaL$ and $\cos\thetap$, and  a product of two complementary error functions for $\cos \thetam$, as shown in Fig.~\ref{fig:sidebands} for \Lambdab candidates at $\sqrt{s} = 8$\TeV.

\begin{figure*}[h!tb]
\centering
\includegraphics[width=0.48\textwidth]{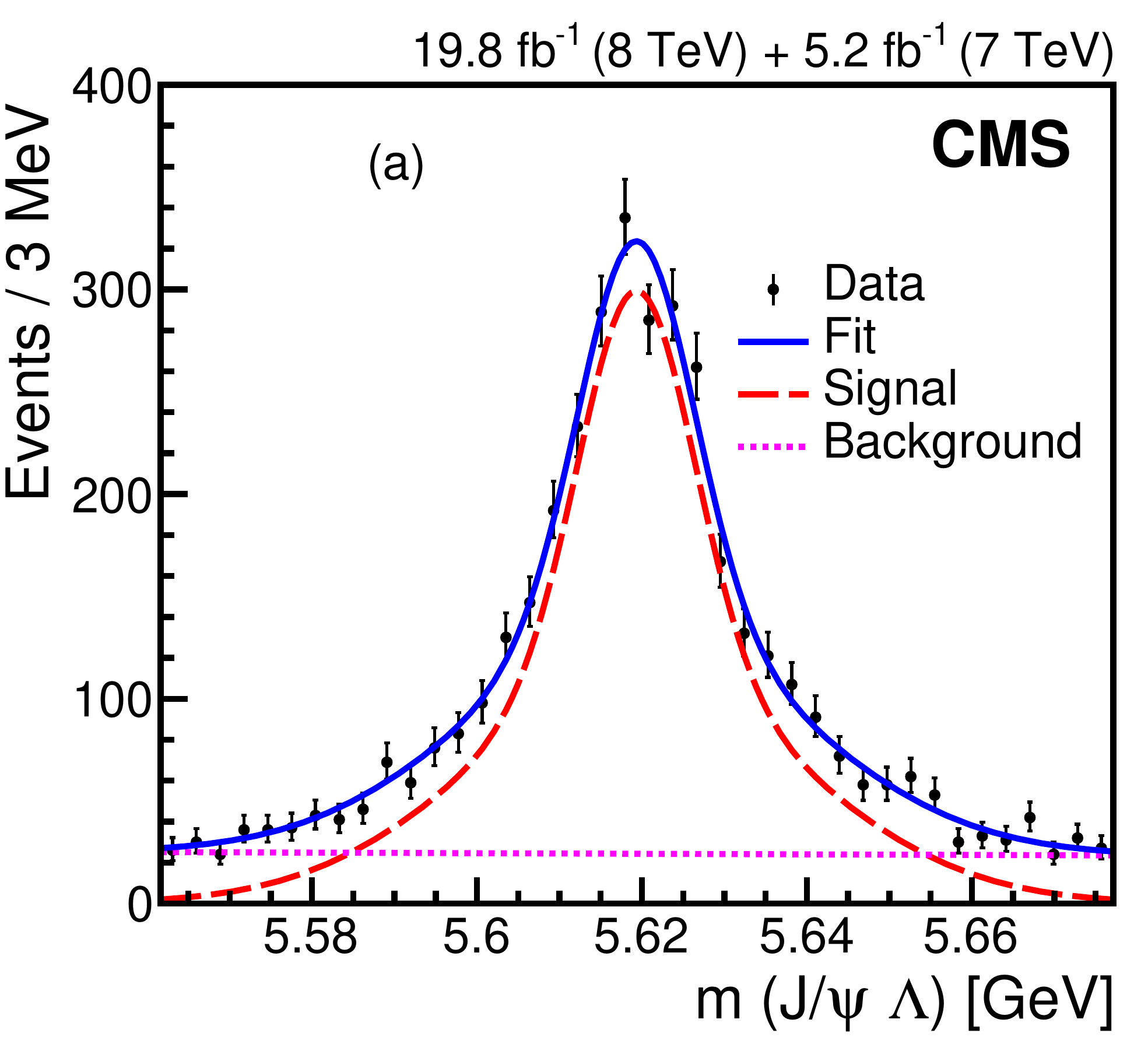}   \includegraphics[width=0.48\textwidth]{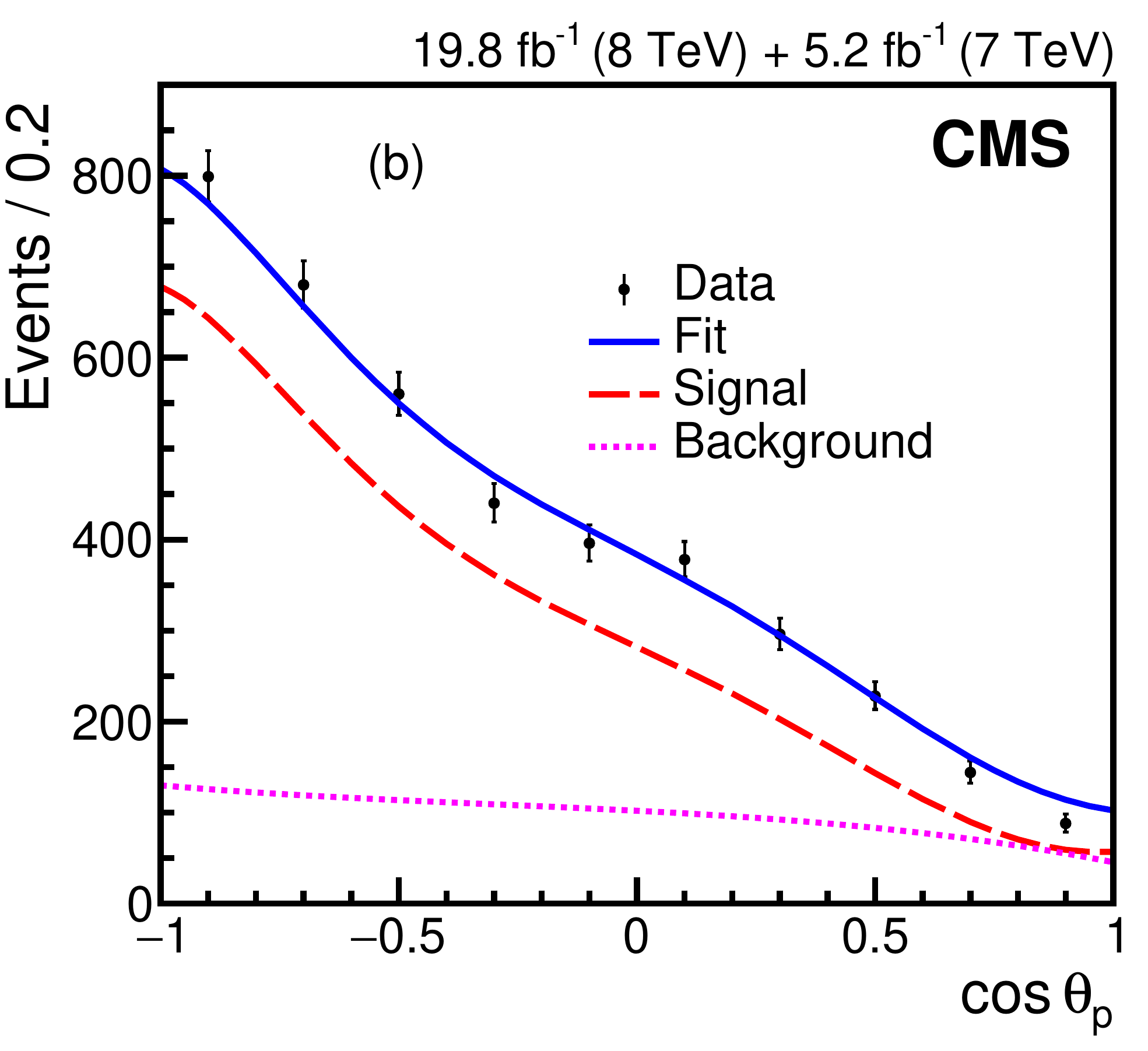}
\includegraphics[width=0.48\textwidth]{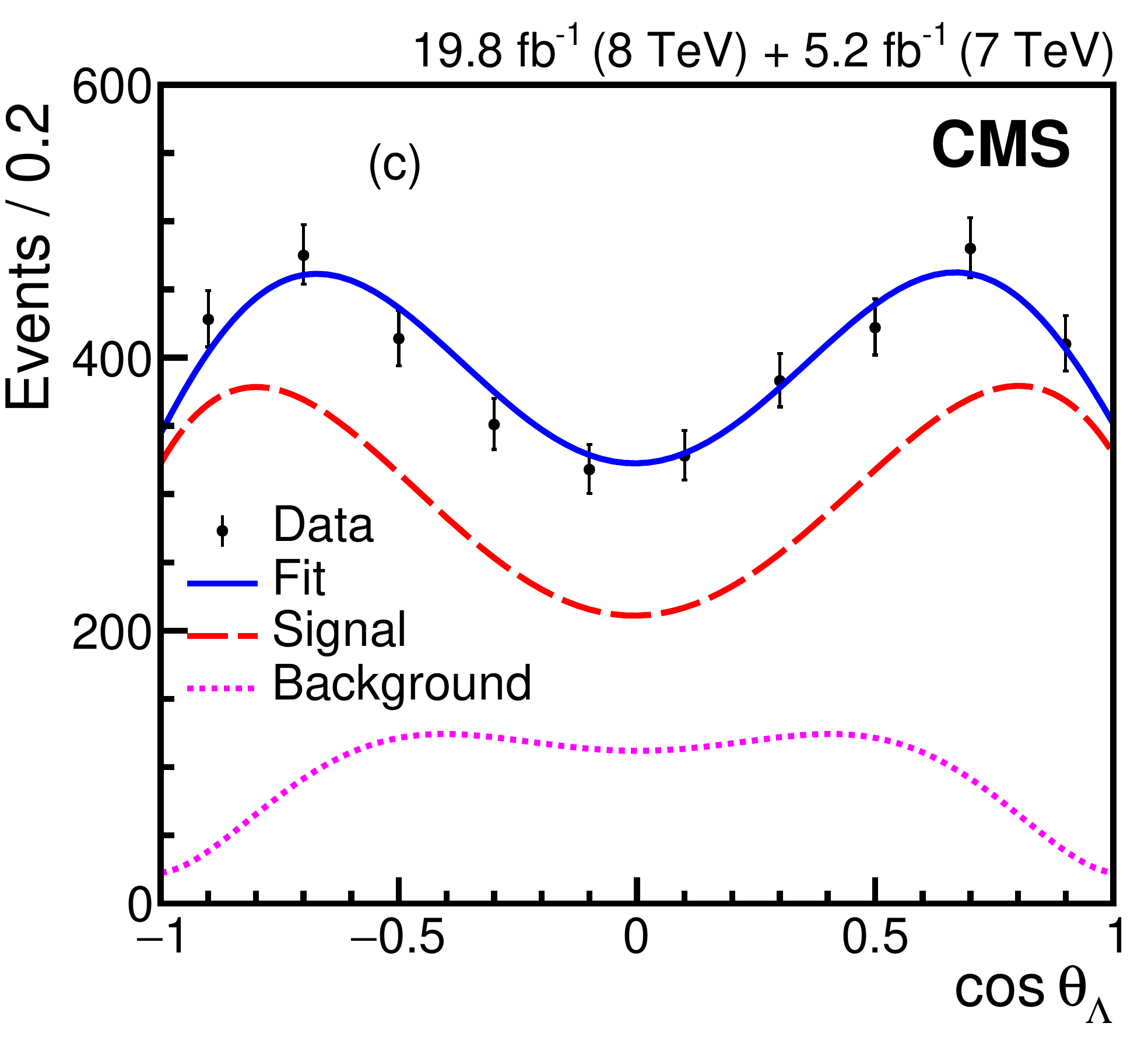} \includegraphics[width=0.48\textwidth]{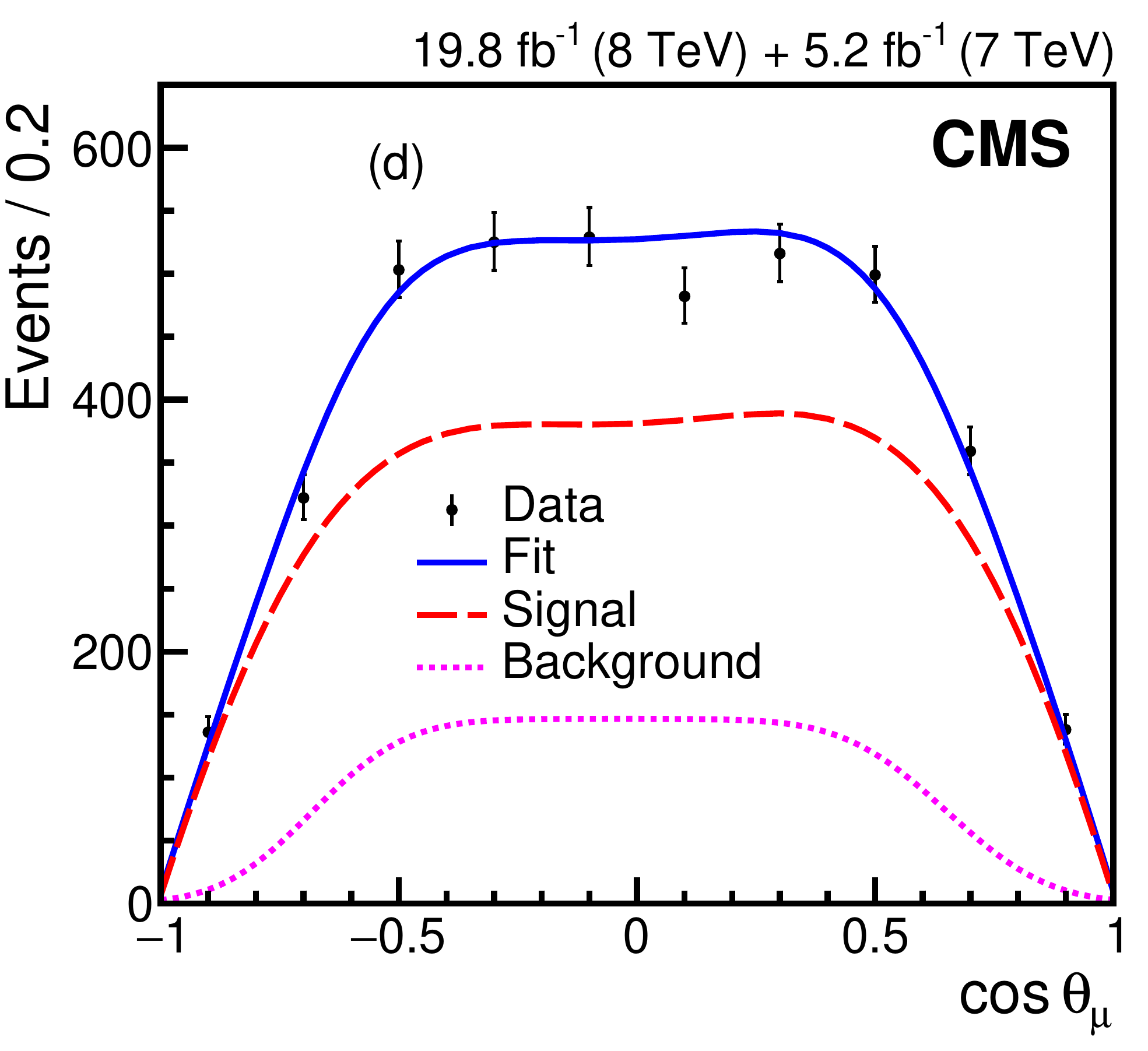}
\caption{Distributions in (a) $\MLb$, (b) $\cos\thetap$, (c) $\cos\thetaL$, and (d) $\cos\thetam$ for \Lambdab candidates in the combined $\sqrt{s}= 7$ and 8\TeV data. The vertical bars on the points are the statistical uncertainties in the data, the solid line shows the result of the fit, and the dashed and dotted lines represent, respectively, the signal and background contributions from the fit.}\label{fig:L_fit35s}
\end{figure*}

\begin{figure*}[h!tb]
\centering
\includegraphics[width=0.48\textwidth]{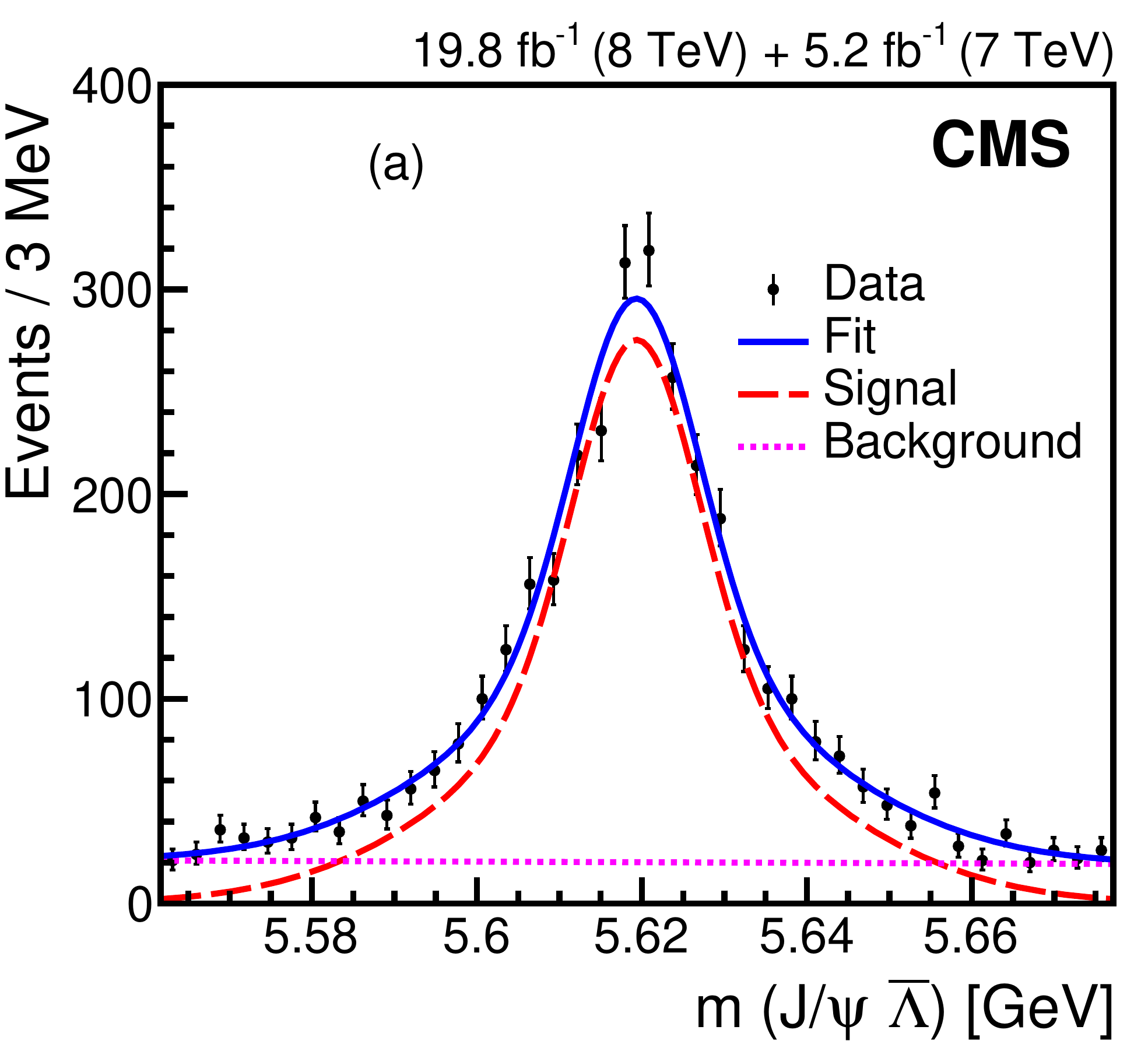}
\includegraphics[width=0.48\textwidth]{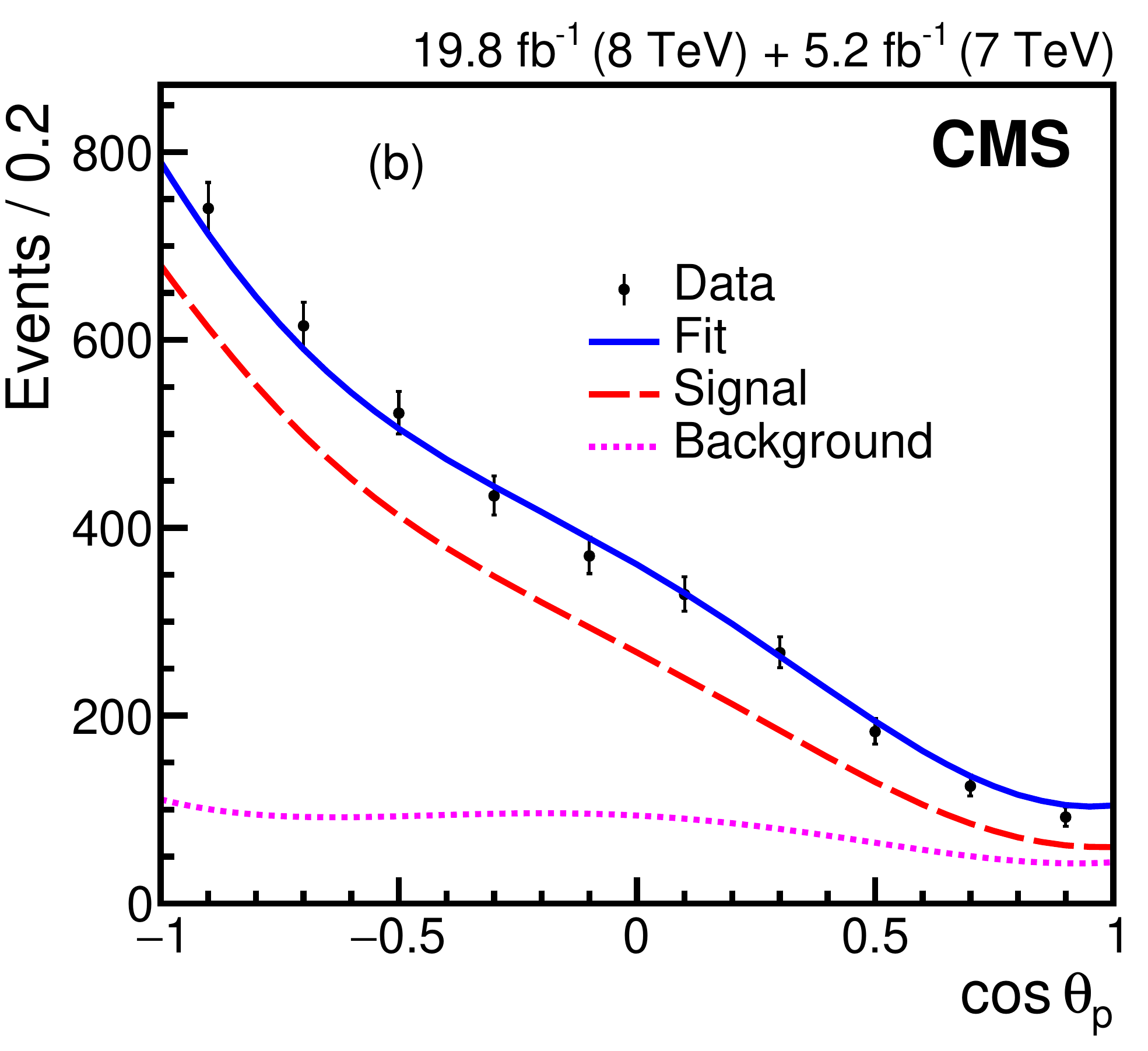}
\includegraphics[width=0.48\textwidth]{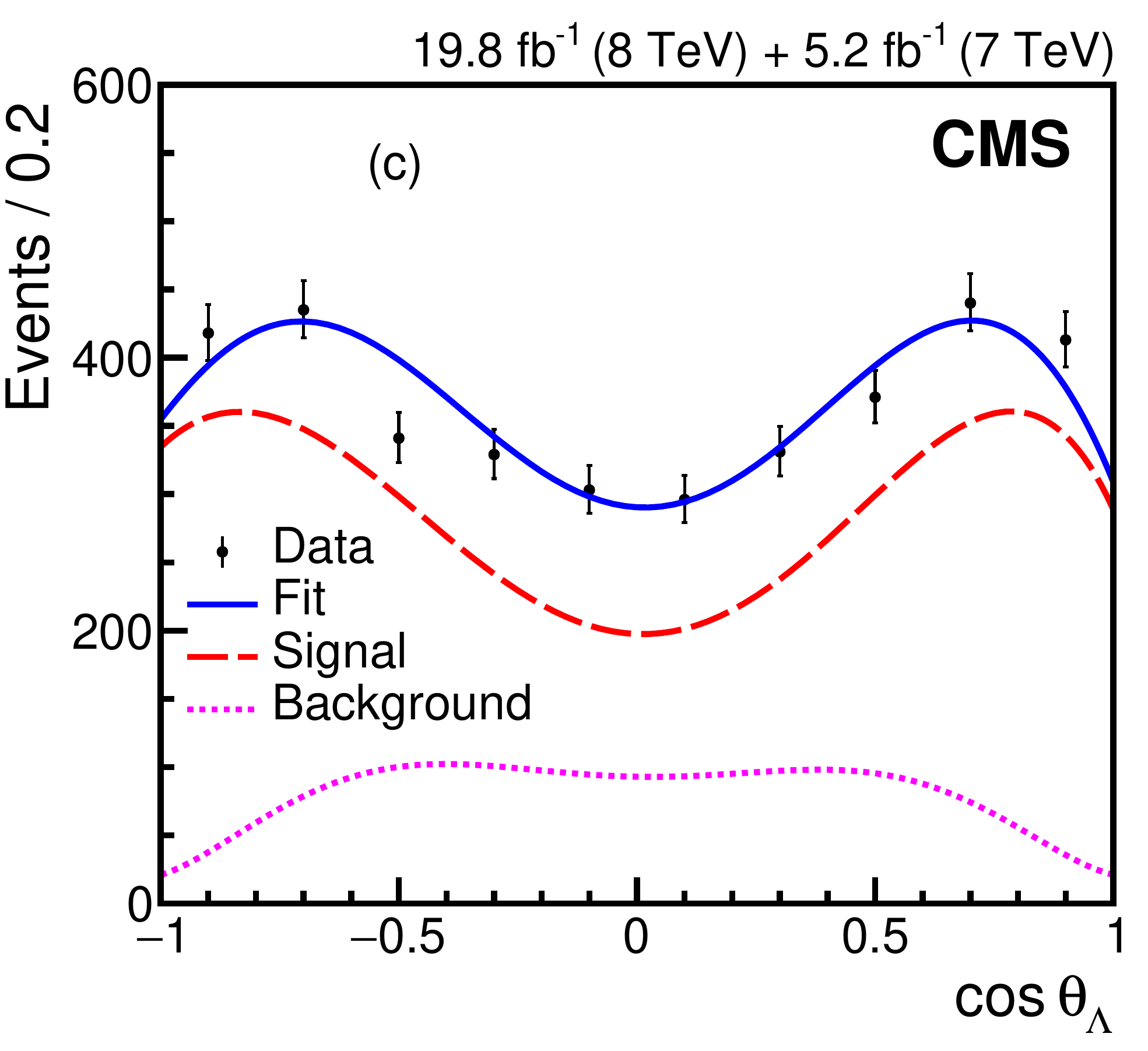}
\includegraphics[width=0.48\textwidth]{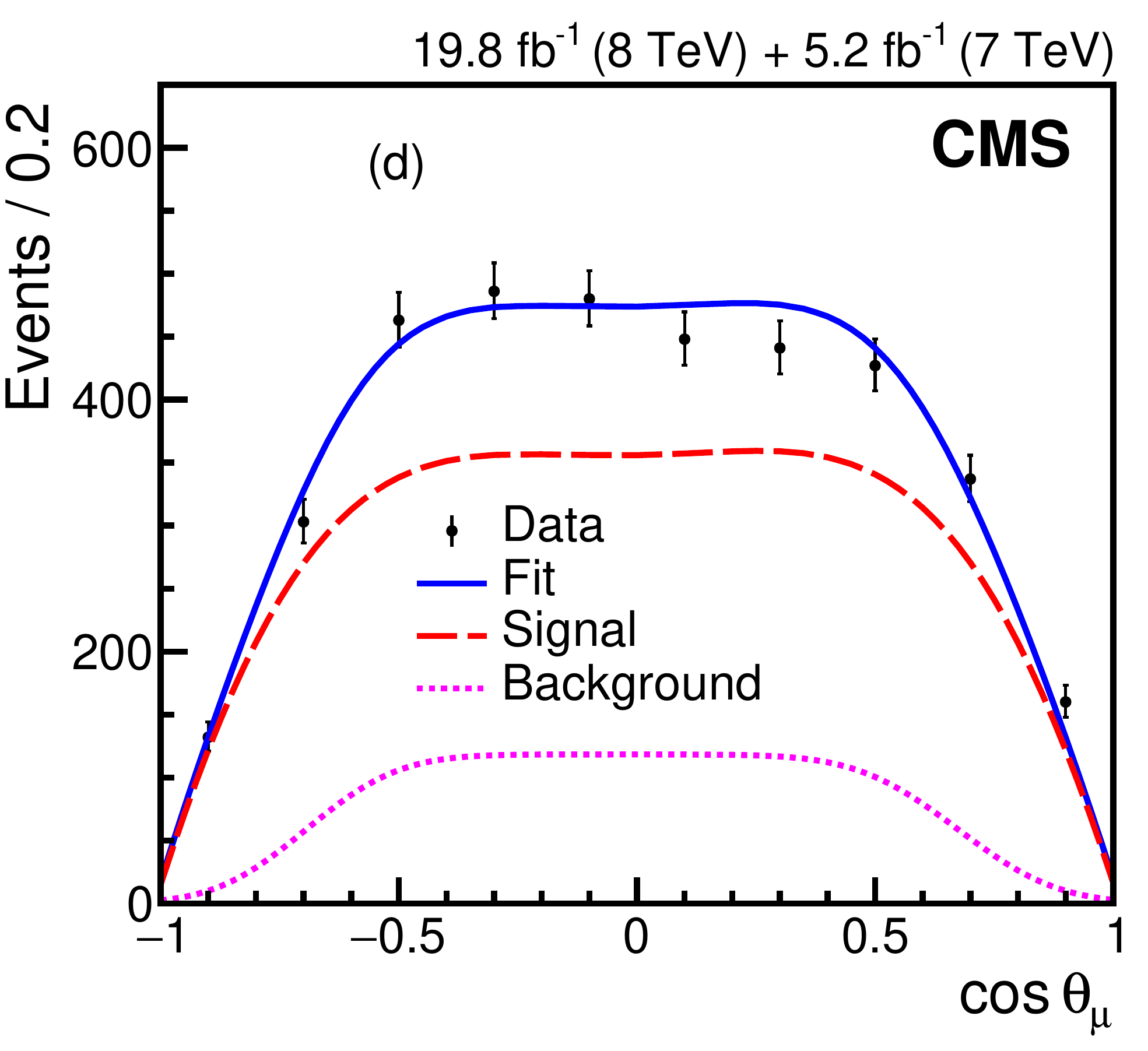}
 \caption{Distributions in (a) $\aMLb$, (b) $\cos\thetap$, (c) $\cos\thetaL$, and (d) $\cos\thetam$ for \aLambdab candidates in the combined $\sqrt{s}= 7$ and 8\TeV data. The vertical bars on the points are the statistical uncertainties in the data, the solid line shows the result of the fit, and the dashed and dotted lines represent, respectively, the signal and background contributions from the fit.}\label{fig:aL_fit35s}
\end{figure*}

The complete likelihood function in Eq.~(\ref{eq:Flikelihood}) is maximized by fitting simultaneously the four data sets for \Lambdab and \aLambdab at $\sqrt{s}= 7$ and 8\TeV, allowing for the extraction of the common parameters $P$,  $\alpha_{1}$, $\alpha_{2}$, and $\gamma_{0}$. The simultaneous fit is performed in the enriched signal mass range within 3.5 SDs of the mean \Lambdab mass. This range contains more than 99.9\% of the signal events, and significantly reduces the number of background events. As a result, the fit is less sensitive to the modeling discussed above. The fit parameters for the background and efficiency distributions are fixed to those found in the previous fits. The signal and background mass parameters are obtained from previous fits to the mass distribution within 10 SDs, and the numbers of signal and backgrounds events are fixed to the propagated values in the signal mass region. The resulting fit values of the polarization and the three angular decay parameters are
\begin{equation*}
\begin{aligned}
  P &= 0.00 \pm 0.06,\\
  \alpha_{1} &=  0.14 \pm 0.14,\\
  \alpha_{2} &= -1.11 \pm 0.04,\\
  \gamma_0 &= -0.27 \pm 0.08,
\end{aligned}\end{equation*}
 where the uncertainties are statistical only. The correlation matrix of the fitted parameters is shown in Table~\ref{table:correlations}. No strong correlations are found among these parameters.
 Translating these values to the squares of the helicity amplitudes, the results are
\begin{equation*}\begin{aligned}
      	\abs{ T_{++}}^{2} &=& 0.05 \pm 0.04,\\
	\abs{ T_{+0}}^{2} &=& -0.10 \pm 0.04,\\
  	\abs{ T_{-0}}^{2} &=& 0.51 \pm 0.03,\\
  	\abs{ T_{--}}^{2} &=& 0.52 \pm 0.04,
\end{aligned}\end{equation*}
where the uncertainties are statistical only. The projections of the fit are shown in Figs.~\ref{fig:L_fit35s} and~\ref{fig:aL_fit35s}  for \Lambdab and \aLambdab, respectively, using the combined data at $\sqrt{s} = 7$ and 8\TeV.

\begin{table} [h!t]
\centering
\topcaption{Correlation matrix for the fitted parameters.}\label{table:correlations}
\begin{scotch}{ccccc}
   Parameter 	& $P$ 	&$\alpha_1$ & $\alpha_2$ & $\gamma_{0}$ \\ \hline
   $P$       	&  1  	&$-0.039$ & $-0.029$ & $0.116$ 	\\
   $\alpha_{1}$ &     	& 1    	  & $-0.207$ & $-0.030$	\\
   $\alpha_{2}$ &     	&  	  & 1 	     & $0.285$ 	\\
   $\gamma_{0}$ & 	& 	  &  	     &  1	\\	
\end{scotch}
\end{table}

\section{Systematic uncertainties}
Various sources of systematic uncertainty that affect the measurement of the parameters $P$, $\alpha_1$, $\alpha_2$, and $\gamma_0$ are discussed below.

\paragraph*{Fit bias}
 The bias introduced through the fitting procedure is studied by generating 1000 pseudo-experiments using the measured parameters as inputs. The difference between the input and the mean of the fitted values is taken as the systematic uncertainty.

\paragraph*{Asymmetry parameter $\alpha_{\PgL}$}
 This parameter is varied up and down by its uncertainty and the maximum deviation in the final result for each parameter is taken as the systematic uncertainty.

\paragraph*{Model for the background \MLb distribution}
 An exponential function is used instead of the first-order polynomial in the likelihood fit. The parameter of the exponential and the background yield are varied by their uncertainties. The fit is redone taking into account this variation on the background model for the mass, and the differences between these results and the nominal fit results are taken as the systematic uncertainty for this source.

\paragraph*{Model for the  background angular distributions}
 Alternative parametrizations of the background angular distributions are used to estimate the systematic uncertainty. For $\cos\thetaL$ and $\cos\thetam$ the alternative models comprise a superposition of Gaussian kernels, as implemented in Roofit RooKeysPdf~\cite{RooFit}, while for $\cos\thetap$ the alternative model is an error function. The differences relative to the nominal results are taken as the systematic uncertainties from the modeling of the background angular distributions.

\paragraph*{Model for the signal \MLb distribution}
 We estimate this uncertainty by changing the parameters by their uncertainties, taking into account their correlations. In each sample of \Lambdab and \aLambdab at $\sqrt{s}= 7$ and 8\TeV, we use the parameter of the signal mass model with the largest global correlation and add 1 SD to its nominal value  if the correlation is positive and subtract 1 SD if the correlation is negative. The difference relative to the nominal result is quoted as a systematic uncertainty.

\paragraph*{Angular efficiencies}
 The values of the Chebyshev polynomial coefficients that model the angular dependence of the efficiencies are changed by their uncertainties. The difference relative to the nominal fitted result is taken as the systematic uncertainty.

\paragraph*{Angular resolution}
 We study the systematic uncertainty in the angular resolution of the measured observables $\cos\thetaL$, $\cos\thetap$, and $\cos\thetam$  by first  determining the resolution using simulated events, then taking the difference between the generated (before detector simulation) and reconstructed (fully simulated) distributions of the cosines of the three polar angles, and fitting the resulting distributions to Gaussian functions. Using these models, we generate random numbers that are added to the three polar angles of the events at generation. The difference between the obtained parameters from the likelihood fits using the same events, with and without the added random terms, it is quoted as the systematic uncertainty from the angular resolution.

\paragraph*{Azimuthal angle efficiency}
 Uniform azimuthal efficiencies are assumed throughout the analysis. Besides simplifying the measurement from a five- to a three-dimensional angular analysis, this assumption also reduces the number of angular parameters from 6 to 3. The effect of the non uniformity in the \phip and \phim efficiencies is investigated with 500 pseudo-experiments generated using the five-dimensional angular distribution, multiplied by the polar and azimuthal efficiencies obtained from the full simulation, as well as initializing the 3 extra parameters to values away from the physical boundary. The resulting distributions are then fitted to the nominal three-dimensional angular model. Differences in the mean values of $P$, $\alpha_1$, $\alpha_2$, and $\gamma_0$ relative to the input values (set to the nominal results) are taken as the systematic uncertainties from the impact of the non uniformity of the azimuthal efficiencies.

\paragraph*{Weighting procedure}
 To estimate the uncertainty from the weighting procedure, we vary each weight by its uncertainty and use this as a new weight to correct the efficiencies, then redo the fit with these new values. The differences between the results of this fit and the nominal values are taken as the systematic uncertainty in each parameter.

\paragraph*{Reconstruction bias}
 Possible unaccounted reconstruction biases are also considered. To estimate this systematic uncertainty, we use a simulated event sample with input values of the helicity amplitudes and polarization similar to those observed in data. Then, after reconstruction and selection as in  data, we fit the simulated events and take the differences between the input and fit values for every angular parameter and polarization.  Since we are using the full reconstruction of the simulated events,  we subtract in quadrature the  systematic sources involved in the fit from those observed differences, and finally take the square root of this subtraction as the estimate of the systematic uncertainty component due to reconstruction bias. This systematic uncertainty is by far the largest uncertainty however, it is still smaller or  comparable to the corresponding statistical uncertainty.

 The contributions from the different uncertainty sources are assumed to be independent and the total systematic uncertainty is calculated as the quadrature sum of all uncertainties. The values  of the systematic uncertainties in each parameter from the individual sources and their quadrature sum are given in Table~\ref{table:Systematic}.

\begin{table*}[h!tb]
\topcaption { The sources and values of the systematic uncertainties in each parameter and the total uncertainty. Each value in the table should be multiplied by $10^{-2}$ to obtain the corresponding systematic uncertainty.}\label{table:Systematic}
\centering
\begin{scotch}{  l  c c c c }
Source 			& $P(\times10^{-2})$ & $\alpha_1(\times10^{-2})$ & $\alpha_2 (\times10^{-2})$ & $\gamma_0(\times10^{-2})$\\ \hline
Fit bias 				&0.1  &0.3&0.1&0.2\\
Asymmetry parameter $\alpha_{\PgL}$	&0.4  &0.7&2.0&0.6\\
Background \MLb distribution 		&0.01 &0.5&1.0&0.9\\
Background angular distribution 	&0.4  &0.5&0.9&5.0\\
Signal \MLb distribution 		&0.01 &0.3&1.0&1.0\\
Angular efficiencies 			&0.1  &0.3&3.0&1.0\\
Angular resolution 			&1.0  &0.1&2.6&0.8\\
Azimuthal angle efficiency 		&0.1  &1.0&0.3&0.1\\
Weighting procedure 			&0.1  &1.3&0.4&2.0\\
Reconstruction bias 			&5.7  &9.8&2.0&9.1\\[\cmsTabSkip]
Total (quadrature sum)			&5.8  &10.0&5.1&11.1\\
\end{scotch}
\end{table*}

\section{Summary and conclusions}

 Based on an angular analysis of about 6000 \Lbdecay events collected by the CMS experiment at $\sqrt{s}= 7$ and 8\TeV, a measurement of the \Lambdab polarization $P$, the parity-violating asymmetry parameter in the \Lambdab decay $\alpha_1$, the {\PgL} longitudinal polarization  $\alpha_2$, and  the parameter $\gamma_0$ has been performed.  The obtained values are
\begin{equation*}\begin{aligned}
P &= 0.00 \pm0.06\stat\pm0.06\syst,\\
\alpha_1 &= 0.14 \pm 0.14\stat \pm 0.10\syst,\\
\alpha_2 &= -1.11 \pm 0.04\stat \pm 0.05\syst,\\
\gamma_0 &= -0.27 \pm 0.08\stat \pm 0.11\syst,
\end{aligned}\end{equation*}
 corresponding to the squares of the helicity amplitudes
\begin{equation*}\begin{aligned}
  \vert T_{++}\vert^2 &=& 0.05 \pm 0.04\stat \pm 0.04\syst,\\
  \vert T_{+0}\vert^2 &=& -0.10 \pm 0.04\stat \pm 0.04\syst,\\
  \vert T_{-0}\vert^2 &=& 0.51 \pm 0.03\stat \pm 0.04\syst,\\
  \vert T_{--}\vert^2 &=& 0.52 \pm 0.04\stat \pm 0.04\syst.
\end{aligned}\end{equation*}
The measured \Lambdab polarization value given above is consistent with the LHCb measurement~\cite{LHCb} and theoretical predictions of 0.10~\cite{A7} and 0.20~\cite{A8}. Note that in our notation, $\alpha_1$ is the negative value of $\alpha_b$ referred to in the theory~\cite{A16,A17,A18,A19,A20,Gutsche}, LHCb~\cite{LHCb}, and ATLAS~\cite{ATLAS} papers. To compare with the theoretical predictions and the  other measurements, we use the negative of our measured value of $\alpha_1$. The many theoretical predictions for $-\alpha_1$ include $-0.2$ to $-0.1$  from quark model techniques~\cite{A16,A17,A19,A20,Gutsche}, $-0.17$ to $-0.14$ from perturbative quantum chromodynamics calculations~\cite{A18}, and 0.78 from heavy-quark effective theory~\cite{A8,LEITNER2005435}. The measured value is inconsistent at more than 5 standard deviations with the  heavy-quark effective theory prediction, but is consistent at less than one standard deviation with the other predictions. The presented  measurement of $\alpha_1$ is also consistent  with the measurements $0.05 \pm 0.17\stat\pm 0.07\syst$ and $0.30 \pm 0.16\stat\pm 0.06\syst$ by LHCb~\cite{LHCb} and ATLAS~\cite{ATLAS}, respectively, and with no parity violation at the level of one standard deviation. The measurement of  $\alpha_{2}$, compatible with $-1$, indicates that the positive-helicity states of the {\PgL} coming from the \Lambdab decay are suppressed.

\begin{acknowledgments}
We congratulate our colleagues in the CERN accelerator departments for the excellent performance of the LHC and thank the technical and administrative staffs at CERN and at other CMS institutes for their contributions to the success of the CMS effort. In addition, we gratefully acknowledge the computing centers and personnel of the Worldwide LHC Computing Grid for delivering so effectively the computing infrastructure essential to our analysis. Finally, we acknowledge the enduring support for the construction and operation of the LHC and the CMS detector provided by the following funding agencies: BMWFW and FWF (Austria); FNRS and FWO (Belgium); CNPq, CAPES, FAPERJ, and FAPESP (Brazil); MES (Bulgaria); CERN; CAS, MoST, and NSFC (China); COLCIENCIAS (Colombia); MSES and CSF (Croatia); RPF (Cyprus); SENESCYT (Ecuador); MoER, ERC IUT, and ERDF (Estonia); Academy of Finland, MEC, and HIP (Finland); CEA and CNRS/IN2P3 (France); BMBF, DFG, and HGF (Germany); GSRT (Greece); OTKA and NIH (Hungary); DAE and DST (India); IPM (Iran); SFI (Ireland); INFN (Italy); MSIP and NRF (Republic of Korea); LAS (Lithuania); MOE and UM (Malaysia); BUAP, CINVESTAV, CONACYT, LNS, SEP, and UASLP-FAI (Mexico); MBIE (New Zealand); PAEC (Pakistan); MSHE and NSC (Poland); FCT (Portugal); JINR (Dubna); MON, RosAtom, RAS, RFBR and RAEP (Russia); MESTD (Serbia); SEIDI, CPAN, PCTI and FEDER (Spain); Swiss Funding Agencies (Switzerland); MST (Taipei); ThEPCenter, IPST, STAR, and NSTDA (Thailand); TUBITAK and TAEK (Turkey); NASU and SFFR (Ukraine); STFC (United Kingdom); DOE and NSF (USA).

\hyphenation{Rachada-pisek} Individuals have received support from the Marie-Curie program and the European Research Council and Horizon 2020 Grant, contract No. 675440 (European Union); the Leventis Foundation; the A. P. Sloan Foundation; the Alexander von Humboldt Foundation; the Belgian Federal Science Policy Office; the Fonds pour la Formation \`a la Recherche dans l'Industrie et dans l'Agriculture (FRIA-Belgium); the Agentschap voor Innovatie door Wetenschap en Technologie (IWT-Belgium); the Ministry of Education, Youth and Sports (MEYS) of the Czech Republic; the Council of Science and Industrial Research, India; the HOMING PLUS program of the Foundation for Polish Science, cofinanced from European Union, Regional Development Fund, the Mobility Plus program of the Ministry of Science and Higher Education, the National Science Center (Poland), contracts Harmonia 2014/14/M/ST2/00428, Opus 2014/13/B/ST2/02543, 2014/15/B/ST2/03998, and 2015/ 19 / B /ST2 / 02861, Sonata-bis 2012 / 07 / E / ST2 / 01406; the National Priorities Research Program by Qatar National Research Fund; the program Severo Ochoa del Principado de Asturias; the Thalis and Aristeia programs cofinanced by EU-ESF and the Greek NSRF; the Rachadapisek Sompot Fund for Postdoctoral Fellowship, Chulalongkorn University and the Chulalongkorn Academic into Its 2nd Century Project Advancement Project (Thailand); the Welch Foundation, contract C-1845; and the Weston Havens Foundation (USA).

\end{acknowledgments}
\bibliography{auto_generated}

\cleardoublepage \appendix\section{The CMS Collaboration \label{app:collab}}\begin{sloppypar}\hyphenpenalty=5000\widowpenalty=500\clubpenalty=5000\textbf{Yerevan Physics Institute,  Yerevan,  Armenia}\\*[0pt]
A.M.~Sirunyan, A.~Tumasyan
\vskip\cmsinstskip
\textbf{Institut f\"{u}r Hochenergiephysik,  Wien,  Austria}\\*[0pt]
W.~Adam, F.~Ambrogi, E.~Asilar, T.~Bergauer, J.~Brandstetter, E.~Brondolin, M.~Dragicevic, J.~Er\"{o}, A.~Escalante Del Valle, M.~Flechl, M.~Friedl, R.~Fr\"{u}hwirth\cmsAuthorMark{1}, V.M.~Ghete, J.~Grossmann, J.~Hrubec, M.~Jeitler\cmsAuthorMark{1}, A.~K\"{o}nig, N.~Krammer, I.~Kr\"{a}tschmer, D.~Liko, T.~Madlener, I.~Mikulec, E.~Pree, N.~Rad, H.~Rohringer, J.~Schieck\cmsAuthorMark{1}, R.~Sch\"{o}fbeck, M.~Spanring, D.~Spitzbart, A.~Taurok, W.~Waltenberger, J.~Wittmann, C.-E.~Wulz\cmsAuthorMark{1}, M.~Zarucki
% ...
\vskip\cmsinstskip
% ...
1:~~Also at Vienna University of Technology, Vienna, Austria\\
% ...

\end{sloppypar}
\end{document}